\documentclass[aps,pre,reprint,superscriptaddress,nofootinbib,longbibliography,floatfix]{revtex4-2}

\usepackage[T1]{fontenc}
\usepackage[utf8]{inputenc}
\usepackage{lmodern}
\usepackage{microtype}
\usepackage{amsmath,amssymb,mathtools}
\usepackage{bm}
\usepackage{graphicx}
\usepackage{xcolor}
\usepackage[hidelinks]{hyperref}

\graphicspath{{figures/}}

\newcommand{\be}{\begin{equation}}
\newcommand{\ee}{\end{equation}}
\newcommand{\bea}{\begin{eqnarray}}
\newcommand{\eea}{\end{eqnarray}}

\newcommand{\ddd}{\mathrm{d}}
\newcommand{\eee}{\mathrm{e}}
\newcommand{\avg}[1]{\left\langle #1 \right\rangle}
\newcommand{\ploc}{\mathrm{pl}}
\newcommand{\deloc}{\mathrm{deloc}}
\newcommand{\mtyp}{m_0}

\begin{document}

\title{Partial Localization and Multifractality in Heterogeneous Population Growth with Resetting}

\author{Maximilien Bernard}
\affiliation{Laboratoire de Physique Th\'eorique et Mod\`eles Statistiques, CNRS, Universit\'e Paris-Saclay, 91405 Orsay, France}
\affiliation{Laboratoire de Physique de l'\'Ecole Normale Sup\'erieure, CNRS, ENS and PSL University, Sorbonne Universit\'e, Universit\'e Paris Cit\'e, 75005 Paris, France}

\author{Giuseppe Del Vecchio Del Vecchio}
\affiliation{Laboratoire de Physique de l'\'Ecole Normale Sup\'erieure, CNRS, ENS and PSL University, Sorbonne Universit\'e, Universit\'e Paris Cit\'e, 75005 Paris, France}

\author{Jean-Philippe Bouchaud}
\affiliation{Capital Fund Management, 23 rue de l'Universit\'e, 75007 Paris, France}

\author{Pierre Le Doussal}
\affiliation{Laboratoire de Physique de l'\'Ecole Normale Sup\'erieure, CNRS, ENS and PSL University, Sorbonne Universit\'e, Universit\'e Paris Cit\'e, 75005 Paris, France}

\date{\today}

\begin{abstract}
We study a model of heterogeneous multiplicative growth with mean-field redistribution and independent local stochastic resetting.
The model describes populations or wealth distributed over sites with quenched growth-rate heterogeneity, while resets represent random local extinction or bankruptcy events.
At large redistribution or resetting rate, we find a delocalized phase and we compute the growth rate as well as the stationary distribution of the population of a given site.
We show that reset-free growth episodes generate power-law stationary tails with site-dependent exponents.
{For weak resetting and redistribution, favorable sites can undergo long reset-free episodes and localize. This produces a partially localized phase, in which a finite fraction of the total population is carried by a small set of sites whose identity changes on an $O(1)$ time scale within a subextensive pool of $N^{u^\star}$ favorable candidates, with $0<u^\star<1$.}
We derive the corresponding growth rate from a saddle point analysis, and show that the averaged stationary fractions display multifractal behavior.
\end{abstract}

\maketitle

\textit{Introduction.}
Populations, wealths, firm and city sizes often grow multiplicatively, with changes proportional to their current size. At the same time, migration, mutation, trade, or redistribution transfers population or wealth between different sites or agents. This combination appears in models of wealth and income dynamics, city and firm growth, ecology, evolution, and population dynamics \cite{BouchaudMezard2000,gabaix1999zipf,gabaix2009power,Kardar2020}. Multiplicative growth amplifies heterogeneity, while redistribution tends to homogenize the system. Their competition can lead either to a delocalized state, where many sites carry comparable fractions, or to localization/condensation, where a finite fraction is carried by a few favorable sites. In a mean-field version with quenched heterogeneous growth rates driven by \emph{continuous} multiplicative noise, this mechanism produces a genuinely localized phase in which a \emph{fixed} set of $O(1)$ favorable sites carries a finite fraction of the total population indefinitely \cite{BernardMFRandomG,bernard2026fluctuations,BernardBouchaudLeDoussal2026InPreparation}. As we show below, the nature of the temporal fluctuations qualitatively changes this picture.

Sudden local catastrophes (epidemics, earthquakes, bankruptcies, etc.) provide another form of temporal fluctuations.
They can be modeled by stochastic resetting: in its original formulation, a diffusing particle is returned to the origin at Poisson-distributed times, leading to a nonequilibrium stationary state \cite{EvansMajumdar2011}.
Resetting has since been extended to many stochastic processes and interacting systems, see \cite{EvansMajumdarSchehr2020,NagarGupta2023} for reviews.
In many-particle systems, resets may act globally, on the whole system at once, for instance in spin systems or Brownian particles \cite{MagoniMajumdarSchehr2020,biroli2025resetting,BasuKunduPal2019,KarthikaNagar2020}.
It can also act locally, on each particle independently, e.g. in exclusion processes, zero-range processes, and aggregation models \cite{Grange2020,MironReuveni2021,PelizzolaPrettiZamparo2021,Grange2021Aggregation,Grange2024}.

In the population-growth models we consider here, local resetting corresponds to an on-site catastrophic event: the population on one site is brought back to a small value, while the rest of the system continues to grow and exchange population.
A related additive-growth model with redistribution was studied in Ref.~\cite{Santra2022}, although the additive nature of the growth does not allow for localized phases.

\begin{figure}[t]
\centering
\includegraphics[width=0.99\linewidth]{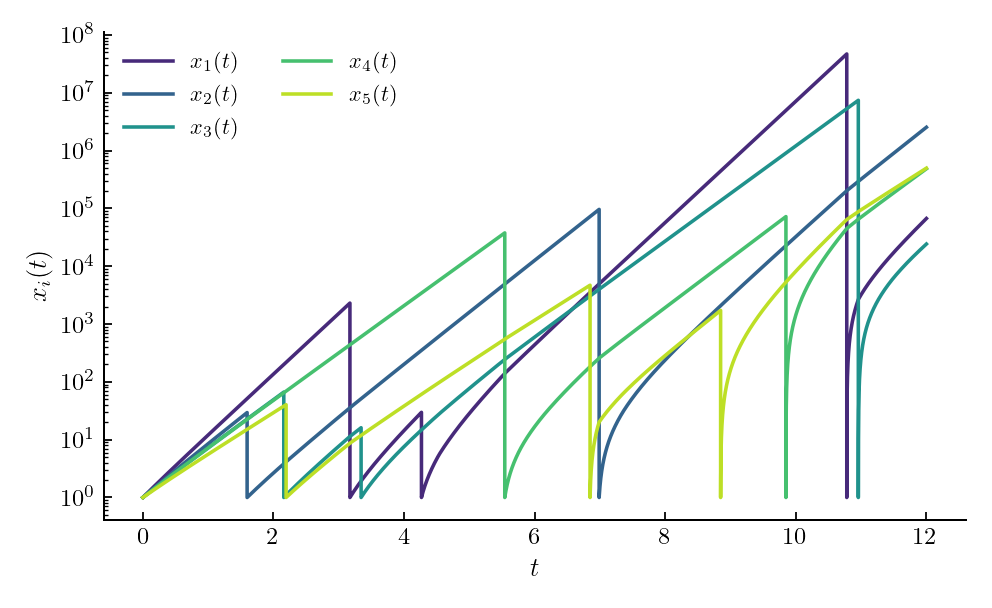}
\caption{
Example trajectories $x_1(t),\ldots,x_5(t)$ for Gaussian heterogeneous growth rates, $\varphi=0.2$, and $r=0.2$.
Resetting interrupts growth episodes through sudden downward jumps.
}
\label{fig:resetting_trajectories}
\end{figure}

\textit{Model.}
We consider $N$ sites with populations $x_i(t)\ge0$, quenched growth rates $m_i$, and mean-field redistribution at rate $\varphi$.
Between resetting events, the evolution of site $i$ is deterministic and given by\footnote{One can add a multiplicative noise term to this equation, as in Refs.~\cite{BouchaudMezard2000, BernardBouchaudLeDoussal2026InPreparation}, with a richer phenomenology; see Appendix S3-A.}
\begin{equation}
\frac{\ddd x_i}{\ddd t}
=
(m_i-\varphi)x_i+\varphi \overline{x},
\qquad
\overline{x}=\frac{1}{N}\sum_{j=1}^N x_j .
\label{eq:model_deterministic_part}
\end{equation}
In addition, site $i$ is reset independently at Poissonian rate $r_i$ to a time-independent value $x_0$,\footnote{One may instead reset to a small fraction of the average population, $x_0 \to \epsilon \overline{x}$.
This slightly modifies the phase diagram, and in particular allows for $\gamma<0$, see Sec.~\ref{sec:frac} of the Supplemental Material.
}
\begin{equation}
x_i\to x_0 .
\label{eq:reset_rule}
\end{equation}
Typical trajectories are shown in Fig.~\ref{fig:resetting_trajectories}. To quantify how the total population grows, we introduce the growth rate of the system as\footnote{This may differ from the annealed growth rate defined through $\langle x(t) \rangle \sim e^{\gamma_a t}$, where $\langle \dots \rangle$ denotes the average over resetting events; see Sec.~\ref{sec:supp_annealed_gamma} of the Supplemental Material.}
\begin{equation}
\gamma=\lim_{t\to\infty}\frac{1}{t}\ln \overline{x}(t).
\label{eq:gamma_def}
\end{equation}
To measure how the population is spread among the $N$ sites, we introduce the fractions
\begin{equation}
p_i(t)=\frac{x_i(t)}{\sum_j x_j(t)},
\qquad
\sum_i p_i(t)=1.
\label{eq:p_def}
\end{equation}

Our main interest is the system once the distribution of the $p_i$'s has reached its steady state for large but finite $N \gg 1$.

\textit{Derivation under a self-averaging assumption.}
Although $\overline{x}(t)$ fluctuates with the resetting noise, in a first stage we neglect these fluctuations and write, at large time, $\overline{x}(t)\simeq x_m\,\eee^{\gamma t}$, with $x_m$ a constant and $\gamma$ to be determined below. In that approximation the sites become effectively independent.
The fractions can then be written as $p_i(t)\simeq x_i/(N x_m\,\eee^{\gamma t})$, and from \eqref{eq:model_deterministic_part} one obtains, between resets,
\begin{equation}
\frac{\ddd p_i}{\ddd t}=a_i p_i+\frac{\varphi}{N},
\qquad
a_i=m_i-\varphi-\gamma,
\label{eq:p_linear}
\end{equation}
while at a resetting event $p_i(t)$ is set to
\begin{equation}
p_0=p_0(t,N)=\frac{x_0\,\eee^{-\gamma t}}{N x_m}.
\label{eq:p0_reset}
\end{equation}
Thus $p_0=O(N^{-1})$ when $\gamma=0$, while $p_0\to0$ exponentially in a growing phase $\gamma>0$. The deterministic solution of \eqref{eq:p_linear} at a time $\tau$ since the last reset is
\begin{equation}
p_i(\tau)=p_i^\star+(p_0-p_i^\star)\,\eee^{a_i \tau},
\qquad
p_i^\star=-\frac{\varphi}{N \,a_i}.
\label{eq:p_solution}
\end{equation}
Let $\tau(p)$ be the time needed to reach the value $p$ after the last reset. Since resetting is Poissonian, the stationary distribution of $p_i$ follows from the renewal formula \cite{EvansMajumdarSchehr2020}
\begin{equation}
P_i(p)=r_i\left|\frac{\ddd \tau}{\ddd p}\right|\eee^{-r_i\tau(p)}.
\label{eq:renewal}
\end{equation}
Inverting Eq.~\eqref{eq:p_solution} gives $\tau(p)=\frac{1}{a_i}
\ln\!\left[
\frac{N p+\varphi/a_i}{N p_0+\varphi/a_i}
\right]$, so that
\begin{align} \label{eq:distpisa}
P_i(p)
&=
|\mu_i|
\left|p_0+\frac{\varphi}{N a_i}\right|^{\mu_i}
\left|p+\frac{\varphi}{N a_i}\right|^{-1-\mu_i},
\nonumber\\
\mu_i
&=
\frac{r_i}{a_i}
=
\frac{r_i}{m_i-\varphi-\gamma},
\end{align}
with support\footnote{With $p_0=0$ for $\gamma>0$. Note that for $x_0>0$ one must have $\gamma \geq 0$.}
\begin{equation}
p
\in
\begin{cases}
\left[p_0,+\infty\right[, \qquad \qquad \qquad \qquad a_i>0,\\[1mm]
\left[\min(p_0,p_i^\star),\max(p_0,p_i^\star)\right],  \quad a_i<0.
\end{cases}
\label{eq:Pi_growth}
\end{equation}
The stationary distribution displays two qualitatively distinct behaviors. If $a_i=m_i-\varphi-\gamma<0$ (and thus $\mu_i<0$), the intrinsic growth rate $m_i-\varphi$ is smaller than the global rate $\gamma$: after a reset $p_i(t)$ relaxes to $p_i^\star=\frac{\varphi}{N(\gamma+\varphi-m_i)}$, and the distribution has bounded support (diverging for $-1<\mu_i<0$, vanishing for $\mu_i<-1$ at the edge).
If $a_i>0$, the site grows faster than the total population between resets; $p_i(t)$ increases exponentially during reset-free intervals whose durations are exponentially distributed, producing the power-law tail in Eq.~\eqref{eq:distpisa} with exponent $\mu_i>0$.
For $p_i=O(1)$, the constraint $\sum_i p_i=1$ imposes an $O(1)$ cutoff whose description goes beyond the self-averaging approximation (see below), so that
\begin{equation}
P_i(p) \sim N^{-\mu_i} p^{-1-\mu_i}, \qquad N^{-1} \ll p \ll 1.
\label{eq:tail_mu}
\end{equation}
The stationary distribution depends on $\gamma$, which is fixed by the normalization $\sum_{i=1}^N \avg{p_i}=1$. If $\mu_i<0$, the distribution has compact support and $\avg{p_i}$ is set by the typical scale $p_i=O(N^{-1})$.
If $\mu_i>1$, a power-law tail is present but its average remains controlled by the same typical scale, $\langle p_i \rangle = O(1/N)$. In both cases, Eq.~\eqref{eq:distpisa} gives
\begin{equation}
\avg{p_i}
=
\frac{\varphi/N+r_i p_0}{\gamma+\varphi+r_i-m_i}.
\label{eq:mean_pi}
\end{equation}
By contrast, for $0<\mu_i<1$, the average is dominated by the $O(1)$ cutoff of the power-law tail and $\langle p_i \rangle \sim N^{-\mu_i}$.\footnote{From \eqref{eq:distpisa} one sees that $\mu_i$ cannot approach $0$.}
The normalization may thus be decomposed as\footnote{We neglect here sites with $\mu_i\simeq 1$, for which both the typical and cutoff contributions may matter.}
\begin{equation}
1=
\sum_{\mu_i>1\ {\rm or}\ \mu_i<0}\avg{p_i}
+
\sum_{0<\mu_i<1}\avg{p_i}.
\label{eq:normalization_decomp}
\end{equation}

We denote the first term $S_{\rm deloc} \equiv \sum_{\mu_i>1\ {\rm or}\ \mu_i<0}\avg{p_i}$ the delocalized contribution, and identify the \emph{delocalized phase} as the region where it dominates, $S_{\rm deloc} \simeq 1$.
This corresponds either to (i) all sites belonging to the first sum in Eq.~\eqref{eq:normalization_decomp}, or (ii) some sites belonging to the second sum, which nonetheless remains negligible as $N\to\infty$.
The normalization $S_{\rm deloc} \simeq 1$ then reads
\begin{equation} \label{selfconsistentdeloc}
1 \simeq  \frac{1}{N}\sum_{\mu_i>1\ {\rm or}\ \mu_i<0} \frac{\varphi+r_i N p_0}{\gamma+\varphi+r_i-m_i}.
\end{equation}
There are two cases: either (i) $\gamma=0$, and \eqref{selfconsistentdeloc} yields an equation for $p_0$ from which $x_m$ is determined; or (ii) $\gamma>0$, determined as the largest root of \eqref{selfconsistentdeloc} with $p_0=0$, which can be written
\begin{equation} \label{eq:eqgammadelocmui}
1 \simeq \frac{1}{N} \sum_{\mu_i>1\ {\rm or}\ \mu_i<0} \frac{\varphi}{r_i(1-\frac{1}{\mu_i})}.
\end{equation}

When this sum includes all sites, \eqref{selfconsistentdeloc} (with $p_0=0$) becomes identical to the equation obtained in \cite{BernardMFRandomG} for the growth rate $\gamma^0$ without resetting ($r_i=0$), up to the shift $m_i\to m_i-r_i$.
In the homogeneous case $m_i=m$, $r_i=r$, Eq.~\eqref{selfconsistentdeloc} gives $\gamma=m-r$, valid as long as $m>r$. If $m<r$, there is no growing solution and, for $x_0>0$, one has $\gamma=0$ with a stationary distribution given by \eqref{eq:distpisa} with $\mu_i=\mu=r/(m-\varphi)$ and $a_i=m-\varphi$. In the heterogeneous case with $r_i=r$ and no $\mu_i\in[0,1]$, the growth rate is $\gamma=\max(0, \gamma^0-r)$, with $\gamma^0$ determined in \cite{BernardMFRandomG}.\footnote{The condition that no $\mu_i\in[0,1]$ corresponds to the delocalized phase $\gamma^0=\gamma^0_{\rm deloc}$ in \cite{BernardMFRandomG}.}

In the general heterogeneous case, the phases and $\gamma$ depend on the distributions of $m_i$ and $r_i$. While populations or wealths can be very large, growth rates are typically finite; we focus on a generalized-Gaussian class. To keep the largest growth rates $O(1)$ as $N\to\infty$, the variance must scale with $N$:
\begin{equation}
\rho_N(m)=\frac{1}{Z_{b,N}}
\exp\!\left[-\left(\frac{(m-\mtyp)^2}{2\Sigma_N^2}\right)^b\right],
\qquad b>0,
\label{eq:generalized-gaussian}
\end{equation}
with $Z_{b,N}$ a normalization and
\begin{equation}
\Sigma_N=\frac{\Sigma_0}{\sqrt{2(\ln N)^{1/b}}},
\label{eq:SigmaN}
\end{equation}
so that $\max_i m_i\to \mtyp+\Sigma_0$ as $N\to\infty$.
The case $b=1$ is Gaussian, while $b=1/2$ corresponds to exponential tails. Most sites have $m_i\simeq \mtyp$, whereas sites near the upper edge $\mtyp+\Sigma_0$ are rare.

Alternatively, one may consider distributions with compact support, e.g.
\begin{equation} \label{eq:eqmcompact}
\rho(m)=\frac{\psi}{ m_>} \left(1 - \frac{m}{m_>}\right)^{\psi -1}
\qquad\text{for } 0<m<m_>,
\end{equation}
which behave differently near the upper edge and hence have different localization properties.
We analyze the generalized-Gaussian class in the main text and discuss the compact-support case in Sec.~\ref{sec:supp_compact_support} of the Supplemental Material.
For simplicity, we take a constant resetting rate $r_i=r$; the results below generalize to heterogeneous rates.

\textit{Delocalized phase.}
Since most $m_i$ are concentrated around $\mtyp$, Eq.~\eqref{selfconsistentdeloc} gives, for $\gamma>0$,
\begin{equation}
1 \simeq S_{\rm deloc} =
\frac{\varphi}{\gamma+\varphi+r-\mtyp}
+O(\Sigma_N^2).
\end{equation}
At leading order, $\gamma=\gamma_{\rm deloc}=\mtyp-r$, with corrections of order $\Sigma_N^2\sim(\ln N)^{-1/b}$. For these sites the tail exponent is $\mu_i=\frac{r}{r-(m_0-m_i)-\varphi}\simeq \frac{r}{r-\varphi}$, consistent with $\mu_i>1$ or $\mu_i<0$. This holds for $\mtyp>r$; for $\mtyp<r$, one has $\gamma_{\rm deloc}=0$ with a site-dependent stationary distribution \eqref{eq:distpisa}. In summary,
\begin{equation}\label{eq:predgaussdeloc}
\gamma_{\rm deloc}=\max(0,\mtyp-r).
\end{equation}

\begin{figure}[t]
\centering
\includegraphics[width=0.99\linewidth]{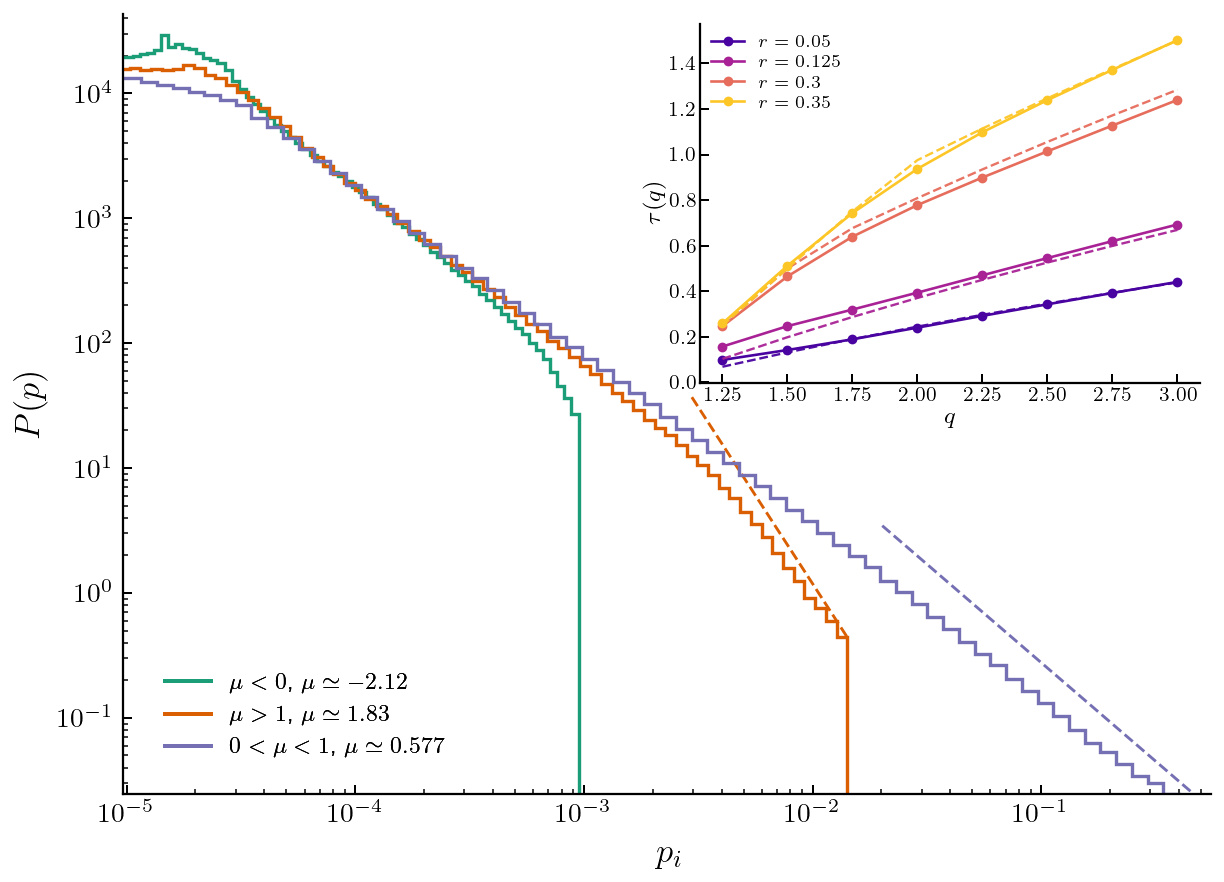}
\caption{
Main: Distribution of the fractions $p_i$ for three sites in the partially localized phase, $\varphi=0.2$, $r=0.1$, $\Sigma_0=1$, $N=2^{15}$, illustrating $\mu_i<0$, $\mu_i>1$, and $0<\mu_i<1$. The self-averaging exponent $\mu_i=r/(m_i-\varphi-\gamma)$ is plotted (dotted) when $\mu_i>0$; $\gamma$ is estimated from \eqref{eq:gamma_def}.
Inset: Multifractal exponent $\theta(q)$ of $Y_q=\sum_i\avg{p_i}^{\,q}$ for $\varphi=0.3$, $\Sigma_0=1$, $\mtyp=0$, and $r=0.05,0.125$ (partially localized) and $r=0.3,0.35$ (delocalized). Dotted: numerical estimates; dashed: Eq.~\eqref{eq:tau_q_saddle}.
}
\label{fig:tau_multifractal}
\end{figure}

\textit{Partially localized phase.}
We now return to the second sum in Eq.~\eqref{eq:normalization_decomp}, $S_{\rm p.l.}\equiv \sum_{0<\mu_i<1}\avg{p_i}$.
When this contribution is $O(1)$, broad-tail sites contribute to the normalization and the fluctuations of $\bar x(t)$ are no longer negligible.
In a first stage we still neglect them, which yields a simple estimate of the tail exponent and the correct phase boundaries; their effect is discussed afterwards.
Within this approximation, once $p_i$ is large enough and $a_i>0$, the multiplicative term $a_i p_i$ dominates the exchange term, so the time to reach $p_i=p$ is
\begin{equation}
\tau(p)\simeq \frac{\ln(Np)}{m_i-\varphi-\gamma}.
\label{eq:tail_time_beyond_sa}
\end{equation}
Since a reset-free interval $\tau=\tau(p)$ occurs with probability $\sim e^{-r\tau}$, this gives, for $N^{-1}\ll p\ll1$,
\begin{equation}
P_i(p)\sim N^{-\mu_i}p^{-1-\mu_i},
\qquad
\mu_i\simeq \frac{r}{m_i-\varphi-\gamma}.
\label{eq:tail_mu_beyond_sa}
\end{equation}
In the main panel of Fig.~\ref{fig:tau_multifractal} we verify that $P_i(p)$ has the predicted power-law tail.
For a site with $0<\mu_i<1$, the average is dominated by the $O(1)$ cutoff,
\begin{equation}
\avg{p_i}\sim N^{-\mu_i}.
\label{eq:mean_pi_broad}
\end{equation}
To simplify, we take the $m_i$ deterministic, ordered $m_1>m_2>\dots$, and given by their quantiles $\int_{m_i}^{+\infty}\rho_N(u)\,du=i/N$. For $i\sim N^u$ with $0<u<1$,
\begin{equation}
m_i\simeq m(u)=\mtyp+\Sigma_0(1-u)^{1/(2b)}.
\label{eq:m_of_u}
\end{equation}
The sites with $0<\mu_i<1$ then contribute
\begin{align}
S_{\rm p.l.}
&\sim
\sum_{0<\mu_i<1}N^{-\mu_i}
\nonumber\\
&\simeq
\ln N\int_0^{u_c}
\ddd u\,\exp\!\left[\ln N\,(u-\mu(u))\right],
\label{eq:SN_saddle}
\end{align}
where
\begin{equation}
\mu(u)=\frac{r}{m(u)-\varphi-\gamma}
\label{eq:g_def}
\end{equation}
is increasing on $u\in[0,1]$, and $u_c$ is defined by $\mu(u_c)=1$. The factor $N^u$ counts the sites at rank $i\sim N^u$, while $N^{-\mu(u)}$ is the probability for such a site to be localized ($p_i=O(1)$). At large $N$, $S_{\rm p.l.}$ is controlled by
\begin{equation}
G(\gamma)=\max_{0\le u\le u_c} \big[u-\mu(u)\big].
\label{eq:G_gamma_def}
\end{equation}
If $G(\gamma)<0$ the broad-tail contribution vanishes; if $G(\gamma)>0$, $S_{\rm p.l.}$ grows with $N$, inconsistent with $\sum_i \langle p_i\rangle=1$.
The partially localized estimate is thus the value for which $G(\gamma_{\rm pl})=0$. Denoting $u^\star$ the maximizer, this requires
\begin{equation}
u^\star-\mu(u^\star)=0,
\qquad
1-\mu'(u^\star)=0
\label{eq:saddle_conditions}
\end{equation}
simultaneously,\footnote{$u^\star=u_c$ is never reached, as it would imply $u^\star=\mu(u_c)=1$, hence $\gamma_{\rm pl}=\mtyp-\varphi-r<\gamma_{\rm deloc}=\mtyp-r$.} giving
\begin{equation}
\gamma_{\rm pl}=m(u^\star)-\varphi-\frac{r}{u^\star}, \qquad r=-m'(u^\star)\,(u^\star)^2.
\label{eq:gamma_pl_general}
\end{equation}
For Eq.~\eqref{eq:m_of_u}, the second equation reads\footnote{For $b<1/2$ the r.h.s. reaches a maximum at $u^\star=4b/(1+2b)$, so the phase can exist only for $r<r_c$.}
\begin{equation}
r=
\frac{\Sigma_0}{2b}(u^\star)^2
(1-u^\star)^{\frac{1}{2b}-1}.
\label{eq:r_u_star}
\end{equation}
The true growth rate is the maximum of the two estimates,
\begin{equation}
\gamma=\max(\gamma_{\rm deloc},\gamma_{\rm pl}).
\label{eq:gamma_max_selection}
\end{equation}
Indeed, $1=S_{\rm deloc}(\gamma)+S_{\rm p.l.}(\gamma)$ with both terms decreasing in $\gamma$, so the solution is unique. If $\gamma_{\rm deloc}>\gamma_{\rm pl}$, then $G(\gamma_{\rm deloc})<0$, $S_{\rm p.l.}\to0$, and the delocalized estimate is consistent; if $\gamma_{\rm deloc}<\gamma_{\rm pl}$, the p.l. contribution would exceed $O(1)$ and the only solution is $\gamma=\gamma_{\rm pl}$. The transition line is $\gamma_{\rm pl}=\gamma_{\rm deloc}$.

The behavior of $S_{\rm p.l.}$ thus defines two phases. When $S_{\rm p.l.}\to0$, the broad-tail sites are irrelevant at large $N$ and $\gamma$ is set by sites with $\langle p_i \rangle=O(1/N)$, Eq.~\eqref{selfconsistentdeloc}; the probability that any site carries $p_i=O(1)$ vanishes, defining the \emph{delocalized phase}.
When $S_{\rm p.l.}=O(1)$, rare reset-free episodes let a site carry a finite fraction of the population. The saddle in \eqref{eq:SN_saddle} selects sites with $m_i\simeq m(u^\star)$, for which $\mu(u^\star)=u^\star$: each carries $p_i\sim O(1)$ with probability $N^{-u^\star}$, but there are $N^{u^\star}$ candidates. The system is therefore localized at a given time, while the dominant site changes within a subextensive pool of $N^{u^\star}$ favorable sites on the $O(1)$ resetting timescale. We call this the \emph{partially localized phase}.

\begin{figure}[t]
\centering
\includegraphics[width=0.98\linewidth]{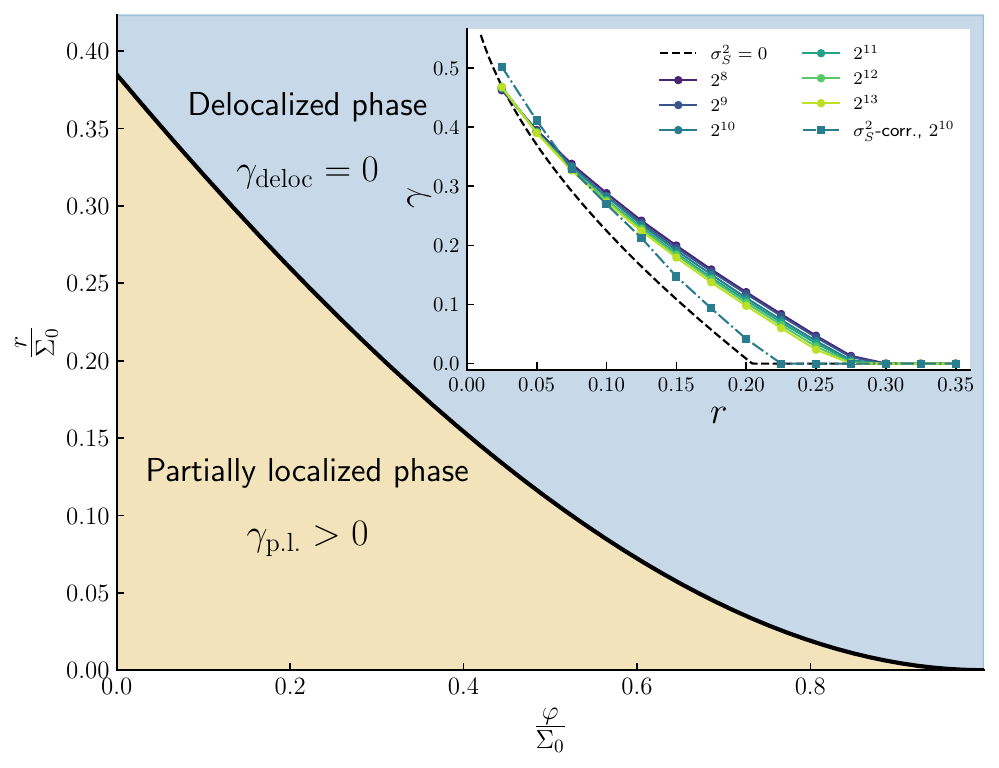}
\caption{Main: Phase diagram in the $(r/\Sigma_0, \varphi/\Sigma_0)$ plane for Gaussian $m_i$. The transition follows from Eq.~\eqref{eq:gamma_max_selection}, $\gamma_{\rm pl}=\gamma_{\rm deloc}$.
Inset: Numerical estimates $\gamma_{\rm est}(N,r)$ for $\varphi=0.3$, $\Sigma_0=1$, $\mtyp=0$, compared with the $N=\infty$ prediction \eqref{eq:predgaussdeloc},\eqref{eq:gamma_pl_general},\eqref{eq:gamma_max_selection} (black dashed). Finite-size effects are of order $1/\ln N$ but their direction is consistent. Square points include the fluctuation term $\sigma_S^2$, estimated numerically.
}
\label{fig:phase_and_growth}
\end{figure}

The above uses the self-averaging approximation. The fluctuations of $\ln\bar x(t)$, encoded in their asymptotic variance $\sigma_S^2$, can be included; the equations no longer close and $\sigma_S^2$ is evaluated numerically (see End Matter). This yields a corrected tail exponent \eqref{eq:mu_sigma} and hence a corrected $\gamma_{\rm pl}$, Eqs.~\eqref{eq:gamma_pl_sigma}, \eqref{eq:ustar_sigma}. Since $\sigma_S$ vanishes at the transition, the phase diagram in Fig.~\ref{fig:phase_and_growth} is unchanged. The correction slightly decreases $\mu_i$ and slightly increases $\gamma_{\rm pl}$, consistent with Fig.~\ref{fig:tau_multifractal} and the inset of Fig.~\ref{fig:phase_and_growth}. Finite-size effects are strong but their direction matches the analytical prediction, and are fitted in Fig.~\ref{fig:logfit}.

\textit{Multifractality.}
In the partially localized phase and for $q>1$, one has $\langle \sum_i p_i(t)^q \rangle=O(1)$: since for $0<\mu_i<1$ the upper cutoff gives $\avg{p_i^q}\sim \avg{p_i}\sim N^{-\mu_i}$, the averaged moment $\sum_i\avg{p_i^q}$ is controlled by the same saddle as the normalization and is $O(1)$.
By contrast, the moments of the averaged fractions $Y_q:=\sum_i\avg{p_i}^{q}$, which for $\mu_i<1$ measure how the probability to be localized is distributed among sites, display multifractal behavior. Using Eq.~\eqref{eq:mean_pi_broad}, the $0<\mu_i<1$ sites contribute
\begin{equation}
Y_q^{\ploc}
\sim
\sum_{0<\mu_i<1} N^{-q\mu_i}
\sim
\int \ddd u\,
\exp\!\left[\ln N\,h_q(u)\right],
\label{eq:Yq_saddle}
\end{equation}
with
\begin{equation}
h_q(u)=u- q\,\mu(u), \qquad \mu(u) \equiv\frac{r}{m(u)-\varphi-\gamma}.
\label{eq:hq_def}
\end{equation}
The $N$ delocalized sites, with $\langle p_i \rangle\sim 1/N$, contribute $Y_q^{\deloc}\sim N^{1-q}$, so that
\begin{equation}
Y_q\sim N^{-\theta(q)},
\qquad
\theta(q)=
-\max\!\left[
1-q,\,
\max_{0\le u\le u_c}h_q(u)
\right].
\label{eq:tau_q_saddle}
\end{equation}
The inset of Fig.~\ref{fig:tau_multifractal} shows the measured $\theta(q)$ for several reset rates, together with the saddle predictions. In the delocalized phase, higher moments can still be dominated by the saddle if some sites have $\mu_i<1$ (red and yellow curves).

For comparison, a fully delocalized phase gives $\theta(q)=q-1$, while complete localization gives $\theta(q)=0$ for $q>1$.
Finally, $\theta(q)$ is related to the fractal-dimension spectrum $d(\alpha)$ by the Legendre transform $\theta(q)=\min_{\alpha}\left[q\alpha-d(\alpha)\right]$, where the number of sites with $\avg{p_i}\sim N^{-\alpha}$ scales as $N^{d(\alpha)}$. In our parametrization, $d(\mu(u))=u$, together with the delocalized point $d(1)=1$.

\textit{Discussion.}
The nature of the temporal noise, and not merely the average dynamics, controls how the population organizes. With continuous multiplicative noise the same model sustains a genuinely localized phase in which a \emph{fixed} $O(1)$ set of favorable sites carries a finite fraction of the total forever \cite{BernardMFRandomG}. Local resetting removes this possibility: after a reset a site restarts from $p_i=O(1/N)$ and needs a reset-free interval of order $\ln N$ to climb back to an $O(1)$ share, while resets recur on an $O(1)$ timescale, so no fixed finite set can survive. What remains is a partially localized phase in which dominance is real but transient: a subextensive pool of $\sim N^{u^\star}$ near-optimal sites competes, and at any instant a small subset holds an $O(1)$ fraction while the leaders continually turn over. Reset-interrupted growth episodes indeed generate power-law tails with \emph{site-dependent} exponents $\mu_i=r/(m_i-\varphi-\gamma)$, and hence a multifractal spectrum $\theta(q)$.
Such a family of models therefore naturally generates weight multifractality,
similar to the fractal and multifractal eigenstates found in generalized
Rosenzweig--Porter random-matrix ensembles
\cite{Kravtsov2015,BogomolnySieber2018,VonSoostenWarzel2019,cugliandolo2024multifractal,KutlinKhaymovich2024}.

This decoupling of \emph{concentration} from \emph{persistence} is, we believe, the most interesting feature of the model, and a natural one in the context of firm or city growth, where resetting mimics bankruptcy or disruptive obsolescence. A very high level of concentration can coexist with permanent reshuffling of the leaders — the losers of today can be the winners of tomorrow — without any contradiction.
A related and more realistic variant would draw a
fresh growth rate $m_i$ at each resetting event. We expect this to leave the phase
diagram unchanged, amounting only to a relabelling of the changing leaders, while making the turnover of dominant
sites explicit.

A second useful output is a sharp criterion for aggregate growth. In the delocalized phase $\gamma=\max(0,\mtyp-r)$: the system grows only if the \emph{typical} unit outpaces the reset rate. In the partially localized phase, growth can be positive even when the typical unit does not grow, because it is carried by the rare, rotating stars; $\gamma_{\rm pl}$ is then set by the best available units rather than the bulk. The transition therefore separates growth as a bulk property from growth as an extreme-value phenomenon. Resetting acts singularly here: for small $r$, $\gamma_{\rm pl}$ approaches its deterministic value as $\gamma_{\rm pl}\simeq \mtyp+\Sigma_0-\varphi-\sqrt{2\Sigma_0 r/b}$ (Supplemental Material), so an arbitrarily small rate of catastrophes has an outsized, non-analytic effect on the growth carried by dominant units.

The case of partial resetting is also interesting, where $x_i \to f x_i + (1 - f) x_0$, generalising the $f=0$ case considered above and treated in SI3.A. This is a stylized model of inheritance tax (tax rate $1-f$), with Poisson death events. When $f > f_c$ is sufficiently large, the localized phase discussed in \cite{BernardMFRandomG} reappears, meaning that dominant dynasties can persist for ever in that case. Within the self-averaging approximation and Gaussian rates the localized phase exists for $f>f_c = \exp( -\sqrt{\Sigma_0/r})$.

Several further extensions are natural: slow evolution of the $m_i$'s on time scale $\sim \ln N$, correlated or anti-correlated $(m_i,r_i)$ (productive units failing less often or becoming more fragile), the compact-support case, which yields only a marginal partially localized phase ($\mu\to1$, $\gamma=m_>-\varphi-r$, Supplemental Material) and underlines the role of the upper edge of $\rho(m)$. Addition of independent white noises to the local growth rates, $m_i \to m_i + \xi_i(t)$, see Supplemental Material, allows to interpolate with the results of Ref.~\cite{BernardMFRandomG}. The most important extension, in our view, is to consider finite dimensional lattices and investigate whether our scenario generates genuine multifractal fields in that case.
Other interesting directions involve adding logistic term as in some ecology models where power law distribution of fractions can also be dynamically generated \cite{al2026spatiotemporal}.

\begin{acknowledgments}
This research was supported by ANR Grant No. ANR-23-CE30-0020-01 EDIPS.
We acknowledge support in part by grant NSF PHY-2309135 to the Kavli Institute for Theoretical Physics (KITP).
\end{acknowledgments}

\newpage

\section*{End Matter}

\begin{figure*}
\centering
\includegraphics[width=0.49\linewidth]{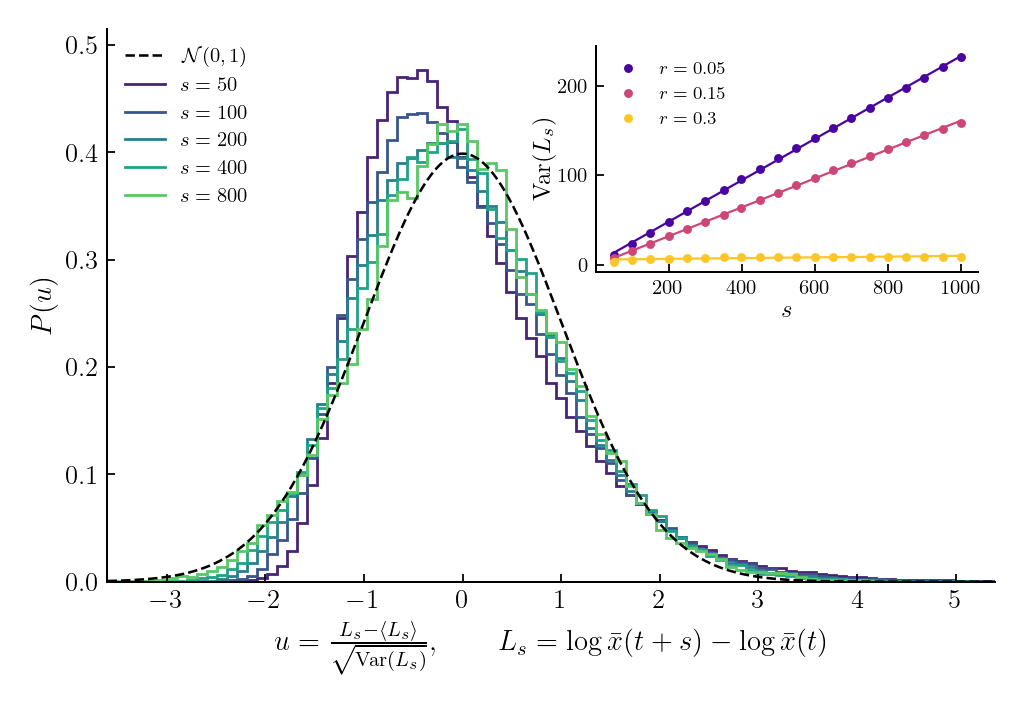}
\includegraphics[width=0.49\linewidth]{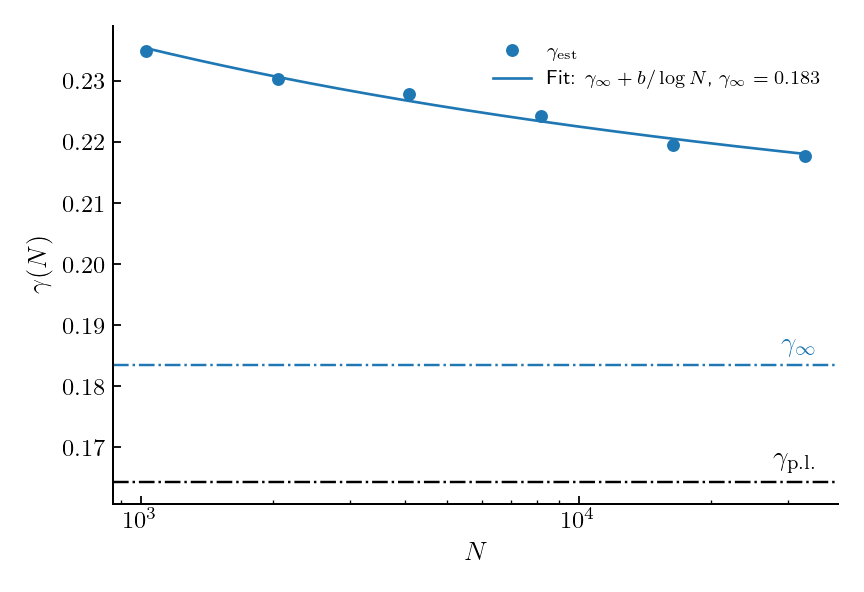}
\caption{Left: Rescaled stationary distribution of $L_s \equiv \ln{\bar x(t+s)}-\ln{\bar x(t)}$ for $\varphi=0.3$, $r=0.125$,$m_0=0,\Sigma_0=1$ (partially localized phase) and $N=2^{11}$. We observe that it converges to a Gaussian at large $s$. Inset: Variance of $L_s$ for several $s$ with a linear fit, confirming the diffusive fluctuations in the partially localized phase ($r=0.05,0.15$), while the fluctuations vanish in the delocalized phase ($r=0.3$). The slope yields the empirical value of $\sigma_S^2$. Right: $\gamma(N)$ for $\varphi=0.3,r=0.125,\Sigma_0=1$ together with a fit of the form $\gamma_\infty+b/\ln{N}$. The estimated $\gamma_\infty \simeq 0.183$ is slightly larger than the predicted value $\gamma_{\rm p.l.}\simeq 0.164$. (Note that the saddle-point estimate seems to suggest the existence of additional $O(\ln{\ln{N}}/\ln{N})$ corrections, which are however difficult to fit).
}
\label{fig:logfit}
\end{figure*}

{\it Tail exponent}. We derive here the correction to the tail exponent taking into account the fluctuations of the total population. From Eq.\eqref{eq:model_deterministic_part} the total population satisfies,
\begin{equation}
\frac{\ddd\ln \bar x}{\ddd t}
=
\sum_j m_jp_j.
\label{eq:endmatter_logX_det}
\end{equation}
while the fractions obey the exact equation
\begin{equation}
\frac{\ddd p_i}{\ddd t}
=
\left(m_i-\varphi-\frac{\ddd\ln \bar x}{\ddd t}\right)p_i
+\frac{\varphi}{N}.
\label{eq:endmatter_exact_pi_log}
\end{equation}
Let $t_0$ be the last resetting time of site $i$. If site $i$ is not reset between $t_0$ and $t$, integration gives
\begin{align}
p_i(t)
&=
p_i(t_0^+)
\exp\!\left[
(m_i-\varphi)(t-t_0)
-\ln\frac{\bar x(t)}{\bar x(t_0^+)}
\right]
\nonumber\\
&\quad+
\frac{\varphi}{N}
\int_{t_0}^t\ddd s\,
\exp\!\left[
(m_i-\varphi)(t-s)
-\ln\frac{\bar x(t)}{\bar x(s)}
\right].
\label{eq:endmatter_pi_logX}
\end{align}
In the partially localized phase, $\gamma>0$ and the first term is negligible.

Moreover, we assume that the fluctuations of $\ddd\ln \bar x/\ddd t$ have an $O(1)$ correlation time set by the resetting time, and that the accumulated fluctuations are Gaussian and diffusive. This is checked numerically in Fig. \ref{fig:logfit} (Left). We therefore write
\begin{equation}
\ln\frac{\bar x(t)}{\bar x(t-s)}
\simeq
\gamma s+\sigma_S B_s,
\label{eq:logx_brownian}
\end{equation}
where $B_s$ is a standard Brownian motion.
Eq.~\eqref{eq:endmatter_pi_logX} becomes
\begin{equation}
p_i(t)
\simeq
\frac{\varphi}{N}
\int_0^\tau \ddd s\,
\exp\!\left[a_i s+\sigma_S B_s\right],
\qquad
\tau=t-t_0 .
\label{eq:endmatter_brownian_integral}
\end{equation}
with $a_i=m_i-\varphi-\gamma$ and $\tau$ is the time since the last reset of site $i$.
The upper tail is controlled by the exponential scale of the last reset-free episode. We therefore write, for the purpose of extracting the exponent,
\begin{equation}
p_i(t)\simeq C_i
\exp\!\left[a_i \tau+\sigma_S B_\tau\right],
\qquad
C_i=O(N^{-1}),
\end{equation}
Here $\tau$ is distributed exponentially with rate $r$, and independent of $B$. Let
\begin{equation}
w=\ln\frac{p_i}{C_i}=a_i \tau+\sigma_S B_\tau.
\end{equation}
Conditioned on $\tau=s$, $w$ is Gaussian with mean $a_i s$ and variance $\sigma_S^2 s$. Averaging over the exponential distribution of $\tau$ gives
the PDF of $w$, $P(w)$ as
\begin{align}
P(w)
&=
\int_0^\infty \ddd s\,
r\eee^{-r s}
\frac{1}{\sqrt{2\pi\sigma_S^2 s}}
\exp\!\left[-\frac{(w-a_i s)^2}{2\sigma_S^2s}\right]
\nonumber\\
&=
\frac{r}{\Delta_i}
\exp\!\left[
\frac{a_i w-\Delta_i |w|}{\sigma_S^2}
\right],
\label{eq:endmatter_fx}
\end{align}
where $\Delta_i=\sqrt{a_i^2+2r\sigma_S^2}$. For $w>0$, $P(w) \sim \exp(-\mu_i w)$, with
\begin{equation}
r
=
a_i\mu_i+\frac{\sigma_S^2}{2}\mu_i^2 .
\label{eq:mu_quadratic_sigma}
\end{equation}
Equivalently,
\begin{equation}
\mu_i
=
\frac{\sqrt{a_i^2+2r\sigma_S^2}-a_i}{\sigma_S^2}
=
\frac{2r}{\sqrt{a_i^2+2r\sigma_S^2}+a_i}.
\label{eq:mu_sigma}
\end{equation}
One recovers Eq.\eqref{eq:distpisa}, $\mu_i=r/a_i$ when $\sigma_S^2\to0$. Since $p_i=C_i\eee^w$, the change of variables gives the upper tail
\begin{equation}
P_i(p)\sim N^{-\mu_i}p^{-1-\mu_i},
\qquad
N^{-1}\ll p\ll 1,
\label{eq:tail_mu_sigma}
\end{equation}
with an $O(1)$ cutoff imposed by normalization.

At fixed $\sigma_S^2$, the partially localized estimate follows from the same saddle calculation as in the main text.
\begin{equation}
S_{\rm p.l.}
\sim
\int \ddd u\, N^{u-\mu(u)} ,
\end{equation}
where $m(u)$ is the rank profile and $\mu(u)$ is given by Eq.~\eqref{eq:mu_sigma}, with $a_i$ replaced by
$m(u)-\varphi-\gamma$. At an interior saddle $u^\star$,$
\mu(u^\star)=u^\star$ and $\mu'(u^\star)=1$ which yields
\begin{equation}
\gamma_{\rm pl}(\sigma_S^2)
=
m(u^\star)-\varphi-\frac{r}{u^\star}
+\frac{\sigma_S^2}{2}u^\star,
\label{eq:gamma_pl_sigma}
\end{equation}
with
\begin{equation}
r
=
-(u^\star)^2m'(u^\star)
-\frac{\sigma_S^2}{2}(u^\star)^2.
\label{eq:ustar_sigma}
\end{equation}
For the generalized-Gaussian distribution $m(u)=\mtyp+\Sigma_0(1-u)^{1/(2b)}$, this becomes
\begin{equation}
r
=
\frac{\Sigma_0}{2b}(u^\star)^2
(1-u^\star)^{\frac{1}{2b}-1}
-\frac{\sigma_S^2}{2}(u^\star)^2.
\label{eq:ustar_sigma_generalized_gaussian}
\end{equation}
These equations reduce to Eqs.~\eqref{eq:gamma_pl_general} and \eqref{eq:r_u_star} for $\sigma_S^2=0$. They are corrected estimates at fixed $\sigma_S^2$, not a closed prediction, because $\sigma_S^2$ is itself determined numerically from the slope of mean square variation, as done in Fig. \ref{fig:logfit}.

{\it Exact expression for the growth rate $\gamma$}.
Eq. \eqref{eq:endmatter_logX_det} yields a useful expression for the  growth exponent. When site $i$ resets, $\bar x\to \bar x(1-p_i+x_0/(N \bar x))$ and therefore
\begin{equation}
\Delta\ln \bar x
=
\ln\left(1-p_i+\frac{x_0}{(N \bar x)}\right).
\end{equation}
Taking the average, in the growing regime where $x_0/(N \bar x)\to0$,
\begin{equation}
\gamma
=
\sum_j m_j\avg{p_j}
+
\sum_j r_j\avg{\ln(1-p_j)}.
\label{eq:endmatter_gamma_log_balance}
\end{equation}
In the delocalized phase, the relevant sites have either $\mu_i<0$ or $\mu_i>1$, so $p_i=O(N^{-1})$ typically. Expanding $\ln(1-p_i)\simeq-p_i$ in Eq.~\eqref{eq:endmatter_gamma_log_balance} gives
\begin{equation}
\gamma
\simeq
\sum_{\mu_i<0 \, \text{or} \, \mu_i>1} (m_i-r_i)\avg{p_i}.
\end{equation}
Together with the expression for $\avg{p_i}$ given in Eq.~\eqref{eq:mean_pi}, one can check that it is equivalent to Eq.~\eqref{selfconsistentdeloc}. In the partially localized phase, Eq.\eqref{eq:endmatter_gamma_log_balance} gives a direct relation between the growth rate $\gamma$ and the marginal distributions $P_i(p)$.

\bibliographystyle{apsrev4-2}
\bibliography{references}

\clearpage
\onecolumngrid
\setcounter{section}{0}
\setcounter{equation}{0}
\renewcommand{\thesection}{S\arabic{section}}
\renewcommand{\theequation}{S\arabic{equation}}

\begin{center}
{\large\bf Supplemental Material}
\end{center}

\section{Annealed Growth Exponent}
\label{sec:supp_annealed_gamma}

The exponent $\gamma$ used in the main text is a typical growth exponent,
\begin{equation}
\gamma=\lim_{t\to\infty}\frac{1}{t}\ln\overline{x}(t).
\end{equation}
One can alternatively define an annealed exponent $\gamma_a$
{
\begin{equation}
\gamma_a =\lim_{t\to\infty}\frac{1}{t}\ln \avg{\overline{x}(t)},
\label{eq:supp_gamma_a_def}
\end{equation}
} where the average is over resetting events for a given set of $m_i$'s. The annealed exponent is sensitive to rare fluctuations and might differ from the typical exponent $\gamma$. {Self-averaging and Jensen’s inequality then imply} $\gamma_a \geq \gamma$.

We first recall the one-site calculation.
Consider
\begin{equation}
\dot x=m x,
\qquad
x\to x_0
\end{equation}
at Poissonian rate $r$.
If no reset has occurred before time $t$, the position is $x_0\eee^{mt}$ and this event has probability $\eee^{-rt}$.
Otherwise, the last reset occurred a time $\tau$ before observation, with probability $r\eee^{-r\tau}$.
The renewal formula { for the PDF of $x$, $P(x,t)$,} is therefore
\begin{equation}
P(x,t)
=
\eee^{-rt}\delta\!\left(x-x_0\eee^{mt}\right)
+r\int_0^t\ddd \tau\,\eee^{-r\tau}
\delta\!\left(x-x_0\eee^{m\tau}\right).
\label{eq:supp_single_site_renewal}
\end{equation}
for $x(0)=x_0$.
The { average} follows directly from Eq.~\eqref{eq:supp_single_site_renewal},
\begin{equation}
\avg{x(t)}
=
x_0\eee^{(m-r)t}
+r x_0\int_0^t\ddd \tau\,\eee^{(m-r)\tau}.
\label{eq:supp_single_site_mean}
\end{equation}
For $m<r$, the {average} reaches a stationary value
\begin{equation}
\avg{x}_\infty=x_0\frac{r}{r-m}.
\end{equation}
For $m>r$, it grows as $\avg{x(t)} {\simeq  \frac{m x_0}{m-r}} \eee^{(m-r)t}$, {so $\gamma_a=m-r$. For $m=r$ it also grows, but as
$\avg{x(t)}=x_0(1+ r t)$, so $\gamma_a=0$}.
\\

We now turn to a $N$-site problem, {with a general growth rate matrix $M$ and where sites reset independently.}
Between resets,
\begin{equation}
\dot x_i=\sum_j M_{ij}x_j,
\label{eq:supp_general_linear}
\end{equation}
and site $i$ is reset to $x_{0i}$ at rate $r_i$.
Let ${\cal R}_k$ be the operator which resets site $k$,
\begin{equation}
({\cal R}_k x)_i=x_i(1-\delta_{ik})+x_{0k}\delta_{ik}.
\label{eq:supp_reset_operator}
\end{equation}
{ The superposition of the independent reset clocks is a Poisson process of rate $\sum_j r_j$.} Conditioning on the last reset event {and the site $k$ where it occurs,} gives the renewal equation
\begin{equation}
P(\vec x,t)
=
\delta\!\left(\vec x-\eee^{Mt}\vec x(0)\right)
\eee^{-\sum_j r_j t}
 +
\sum_k r_k\int_0^t\ddd \tau\,\eee^{-\sum_j r_j\tau}
\int \ddd \vec y\,P(\vec y,t-\tau)
\delta\!\left(\vec x-\eee^{M\tau}{\cal R}_k\vec y\right).
\label{eq:supp_vector_renewal}
\end{equation}
{where we use the vector notation $\vec x=(x_1,\dots,x_N)$.}
Taking the first moment of Eq.~\eqref{eq:supp_vector_renewal} yields
\begin{equation}
\avg{\vec x}_t =
\eee^{-\sum_j r_j t}\eee^{Mt}\vec x(0)
\,+ \,
\sum_k r_k\int_0^t\ddd \tau\,\eee^{-\sum_j r_j\tau}
\eee^{M\tau}\avg{{\cal R}_k\vec x}_{t-\tau}.
\label{eq:supp_mean_renewal}
\end{equation}
{where $\avg{\vec x}_t=\avg{\vec x(t)}= \int d^Nx \vec x P(\vec x,t)$,
and
$\avg{{\cal R}_k\vec x}_{t-\tau}$ means the average of the reset-transformed state.}
Using
\begin{equation}
\sum_k r_k({\cal R}_k x)_i
=x_i\left(\sum_k r_k-r_i\right)+r_i x_{0i},
\label{eq:supp_reset_average}
\end{equation}
and the Laplace transform $\widehat{x}_i(s)=\int_0^{+\infty} dt e^{-s t} \avg{x_i(t)}$ one has
\begin{equation}
\sum_j\left[(s+r_i)\delta_{ij}-M_{ij}\right]\widehat{x}_j(s)
=x_i(0)+\frac{r_i x_{0i}}{s}.
\label{eq:supp_laplace_mean}
\end{equation}
Hence,
\begin{equation}
\widehat{x}_i(s)
=
\sum_j
\left[(sI+R-M)^{-1}\right]_{ij}
\left(x_j(0)+\frac{r_jx_{0j}}{s}\right).
\label{eq:supp_laplace_solution}
\end{equation}
where $R_{ij}=r_i \delta_{ij}$, which upon Laplace inversion
determines $\langle x(t) \rangle$. {Note that \eqref{eq:supp_laplace_mean} is equivalent to the equation
\begin{equation}
\frac{\ddd}{\ddd t}\avg{\vec x(t)}
=
(M-R)\avg{\vec x(t)}
+
R\vec x_0.
\label{eq:supp_mean_dynamics}
\end{equation}
}
Therefore, focusing on a symmetric matrix $M$, the annealed growth rate is fixed by the largest eigenvalue of $M-R$ (provided it is positive):
\begin{equation}
\gamma_a=\max\left(0,\lambda_{\max}\left[M-R \right]\right),
\label{eq:supp_gamma_a_matrix}
\end{equation}
for generic initial conditions. If all eigenvalues of $M-R$
are strictly negative, there is a steady state with $\gamma_a=0$ and $\langle \vec x(t) \rangle \to (R-M)^{-1} R \vec x_0$.
If the largest eigenvalue is zero, then $\gamma_a=0$ but the mean may grow polynomially because of the constant reset source.
The above results are true for arbitrary symmetric matrices $M$, hence for the model {defined on any graph, and in any space dimension.}

For the mean-field case,
\begin{equation}
M_{ij}=(m_i-\varphi)\delta_{ij}+\frac{\varphi}{N}.
\label{eq:supp_M_matrix}
\end{equation}
The eigenvalue problem for $M-R$ is the same as the heterogeneous mean-field problem without resetting, studied in Ref.~\cite{BernardMFRandomG}, after the shift $m_i\to m_i-r_i$.
When the largest eigenvalue of $M-R$ is strictly positive, the
annealed growth exponent $\gamma_a$ is the largest solution of
\begin{equation}
1=
\frac{\varphi}{N}\sum_i
\frac{1}{\gamma_a+\varphi+r_i-m_i} = \frac{1}{N}\sum_i
\frac{\varphi}{r_i( 1- \frac{1}{\mu_i^a})}  \quad, \quad \mu_i^a
=
\frac{r_i}{m_i-\varphi-\gamma_a}
\label{eq:supp_gamma_a_finiteN}
\end{equation}
If the largest eigenvalue of $M-R$ is zero or negative, then
$\gamma_a=0$. When it is strictly positive, the growth rate is thus given by $\gamma^0|_{M \to M-R}$ where $\gamma^0$ is the growth rate in the absence of resetting, i.e. for $r_i=0$, determined in \cite{BernardMFRandomG}. The latter is exact and valid in all the
phases of Ref. \cite{BernardMFRandomG}.
We have introduced the annealed analog of the exponents $\mu_i=\frac{r_i}{m_i-\varphi-\gamma}$ defined in the text, with $\mu_i^a \geq \mu_i$.
When all the $\mu_i>1$ {or $\mu_i<0$}, this equation is the same as Eq.\eqref{eq:mean_pi},\eqref{eq:normalization_decomp} and $\mu_i=\mu_i^a$, $\gamma=\gamma_a$. On the other hand, when the smallest positive $\mu_1 =\min_{\mu_i>0}{\mu_i}$ (which is associated to the largest $m_i-r_i$) becomes smaller than $1$,  $\mu_1^a$ is forced to saturate at $1^+$ as $1/(1-1/\mu_1^a)$ in \eqref{eq:supp_gamma_a_finiteN} starts diverging. In this case, $\mu_1^a>\mu_1$ and $\gamma_a>\gamma$.
Hence, when there exist sites with $0<\mu_i<1$, the annealed exponent is instead dominated by rare reset-free periods, and differs from the typical growth exponent. {We emphasize that} it can happen already in the delocalized phase. {Note that individual terms in the sum
\eqref{eq:supp_gamma_a_finiteN} can be interpreted as annealed fractions, i.e. defining
$p_i^a(t) = \langle x_i(t) \rangle/\sum_j \langle x_j(t) \rangle$, one has
for $\gamma_a>0$, $p_i^a = \lim_{t \to +\infty} p_i^a(t)= \frac{\varphi}{N r_i( 1- \frac{1}{\mu_i^a})}$. The case $\mu_i^a\to 1^+$ can thus be
associated to "annealed localization" i.e. the first moment being dominated by rare reset histories associated with the most favorable site.}

For homogeneous parameters $m_i=m$ and $r_i=r$, Eq.~\eqref{eq:supp_gamma_a_finiteN} gives
\begin{equation}
\gamma_a=\gamma=\max{(0,m-r)}.
\end{equation}
For $m_i$ chosen according to the generalized-Gaussian class and for uniform $r=r_i$, the large-$N$ solution of Eq.~\eqref{eq:supp_gamma_a_finiteN} is given by
\begin{equation}
\gamma_a= \max{(0,m_0-r,m_0+\Sigma_0-r-\varphi)}
\label{eq:supp_gamma_a_edge}
\end{equation}
We have used the result of Ref.~\cite{BernardMFRandomG} for $\gamma^0_{\rm loc}$. Since $\lim_{N \to +\infty} \max_i m_i=m_0+\Sigma_0$, the collective/delocalized and edge/localized branches approach $m_0-r$ and
$m_0+\Sigma_0-r-\varphi$, respectively, leading to \eqref{eq:supp_gamma_a_edge}.

\section{Compact-Support Class}
\label{sec:supp_compact_support}

\begin{figure}
\centering
\includegraphics[width=0.49\linewidth]{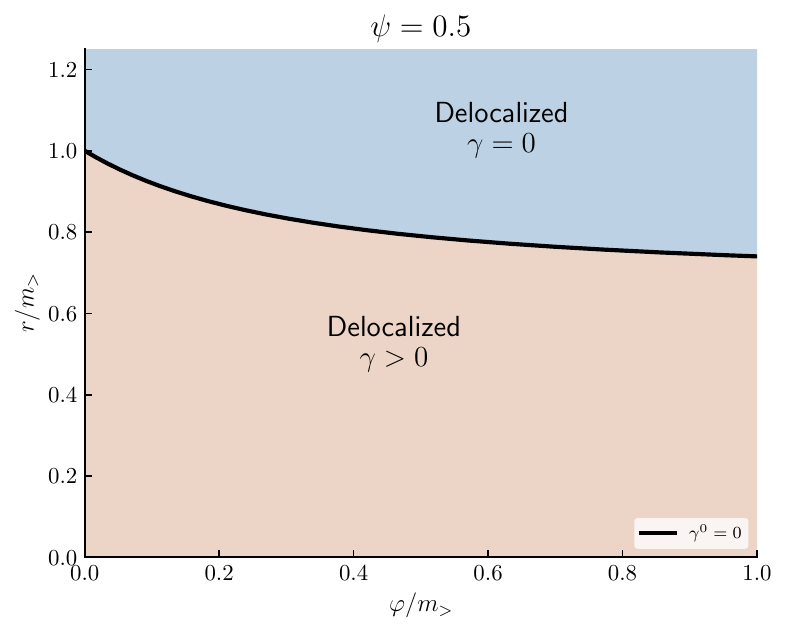}
\includegraphics[width=0.49\linewidth]{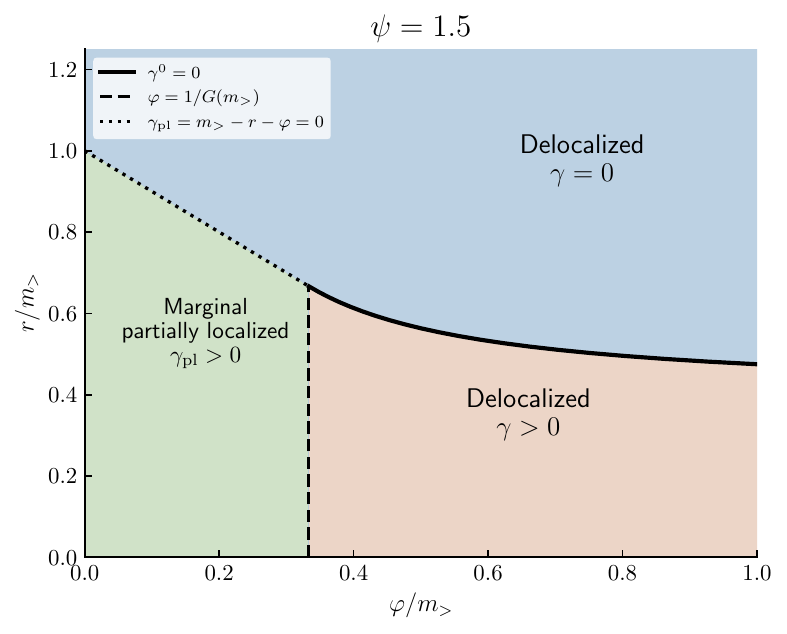}
\caption{
Predicted phase diagram in the ($r/m_>, \varphi/m_>)$ plane for $m_i$ distributed according to Eq.\eqref{eq:supp_compact_density}. Left: $\psi=0.5$, the system is always delocalized, although there are two phases, with and without growth. Right: $\psi=1.5$, for small redistribution, there is an additional marginal partially localized phase
}
\label{fig:phasediagcompact}
\end{figure}

We now estimate the growth rate and the distribution of the $p_i$ when the
$m_i$ are drawn from a distribution with a finite upper edge $m_>$, near which
the density behaves as $\rho(m)\sim \left(m_>-m\right)^{\psi-1}$ with $\psi>0$. In particular, we will study
\begin{equation}
\rho(m)
=
\frac{\psi}{m_>}
\left(1-\frac{m}{m_>}\right)^{\psi-1},
\qquad
0<m<m_>.
\label{eq:supp_compact_density}
\end{equation}
and a uniform resetting rate $r_i=r$.

\subsection{Delocalized phase} \label{Db}

In the delocalized phase, we show below that no site has $
\mu_i=\frac{r}{m_i-\varphi-\gamma}$ in $[0,1]$. For $\gamma>0$, the normalization condition $\sum_{i=1}^N\langle p_i \rangle=1 $ leads to the exact equality
\begin{equation}
1=
\frac{\varphi}{N}\sum_i
\frac{1}{\gamma + r+\varphi-m_i}
\label{eq:supp_compact_deloc_finiteN}
\end{equation}
This equation is identical to the one obtained in Ref.~\cite{BernardMFRandomG} for the growth rate $\gamma^0$ in
the absence of resetting, i.e. for $r=0$, which implies that
in that case
\be  \label{eq:defgammagamma0}
\gamma = \gamma^0 - r.
\ee
which is valid for arbitrary $N$. Note that the condition that no $\mu_i$ is in $[0,1]$
implies that $\gamma > m_i - r - \varphi$ for all $i$, hence from
\eqref{eq:defgammagamma0}, that $\gamma^0 > m_i - \varphi$, for all i.

In the limit of large $N$, it was shown that $\gamma^0= \gamma^0(\varphi)$ is
determined as the solution of the equation
\begin{equation}
\frac{1}{\varphi}=G(\gamma^{0}+\varphi),
\qquad \gamma^{0}+\varphi>m_>, \qquad
G(z)=\int \ddd m\,\frac{\rho(m)}{z-m}.
\label{eq:supp_compact_stieltjes}
\end{equation}
Since we assume here that
$\mu_1 =\min_{\mu_i>0}{\mu_i}>1$, it implies that, in the
large $N$ limit,
$\gamma^0>m_> - \varphi$, i.e. the delocalized branch,
hence there exists a unique solution
to \eqref{eq:supp_compact_stieltjes}. For the model \eqref{eq:supp_compact_density}, one has the explicit expression $G(z)=
\frac{1}{z}
{}_2F_1\!\left(1,1;\psi+1;\frac{m_>}{z}\right)$, as shown in \cite{BernardMFRandomG}.

When the resetting rate is sufficiently large, so that $\gamma^{0}(\varphi)-r$ becomes negative, the system instead has a delocalized stationary state with $\gamma=0$.

The delocalized growth exponent is therefore
\begin{equation}
\gamma_{\rm deloc}=\max\left(\gamma^{0}(\varphi)-r,0\right).
\label{eq:supp_compact_gamma_deloc}
\end{equation}
Hence, within the delocalized phase, there is a transition
at $\varphi=\varphi_c^0$ between
a phase with growth and a phase without growth, where
$\varphi_c^0$ is solution of
\be
\gamma^{0}(\varphi_c^0)=r
\ee
where we recall that $\gamma^{0}(\varphi)$ is a decreasing
function. In
Fig. \ref{fig:phasediagcompact} we show the region
of the phase diagram where this transition occurs.

\subsection{Marginal partially localized branch} \label{Mplb}

We now study the existence and properties of a localized phase. We use again the quantiles of the distribution
For the  density \eqref{eq:supp_compact_density}, they satisfy
\begin{equation}
\frac{i}{N}
=
\left(1-\frac{m_i}{m_>}\right)^\psi,
\qquad
m_i=m_>\left[1-\left(\frac{i}{N}\right)^{1/\psi}\right].
\label{eq:supp_compact_quantiles}
\end{equation}

We recall the tail exponent $
\mu_i=\frac{r}{m_i-\varphi-\gamma}$,
and we now show that one cannot have $\mu_1<1$. We first assume that it is possible and then show a contradiction.
Introducing
\begin{equation}
\Delta=m_>-\varphi-r-\gamma
\end{equation}
one has, with $y=m_>-m$,
\begin{equation}
\mu(y)=\frac{r}{r+\Delta-y}.
\end{equation}
The sites with $0<\mu(y)<1$ have $0<y<\Delta$.
Since the number of sites with $y<m_i<y+dy$ is given by $N \rho(m_>-y) dy$, the partially localized contribution to the normalization, $S_{\rm p.l.} = \sum_{0<\mu_i<1} \langle p_i \rangle$, reads
\begin{equation}
S_{\rm p.l.}
\sim
N\int_0^\Delta \ddd y\,\rho(m_>-y)\,N^{-\mu(y)}.
\end{equation}
Using $\rho(m_>-y)\sim y^{\psi-1}$, this gives
\begin{equation}
S_{\rm p.l.}
\sim
\int_0^\Delta \ddd y\,y^{\psi-1}
\exp\!\left[
\ln N\,\frac{\Delta-y}{r+\Delta-y}
\right].
\label{eq:supp_compact_Spl_integral}
\end{equation}
For fixed $\Delta>0$, the exponent is positive at $y=0$, so $S_{\rm p.l.}$ diverges with $N$.
The normalization therefore forces $\Delta\to0$, i.e.
\begin{equation}
\mu_1\to1,
\qquad
\gamma\to m_>-\varphi-r.
\end{equation}

In other words, the extensive number of sites near the upper edges forces $\mu(m_>)$ to saturate at $1$ and the growth rate is $\gamma=\gamma_{\rm pl} = m_>-\varphi-r$.
We thus denote this phase as marginal partially localized.

\begin{figure}
\centering
\includegraphics[width=0.99\linewidth]{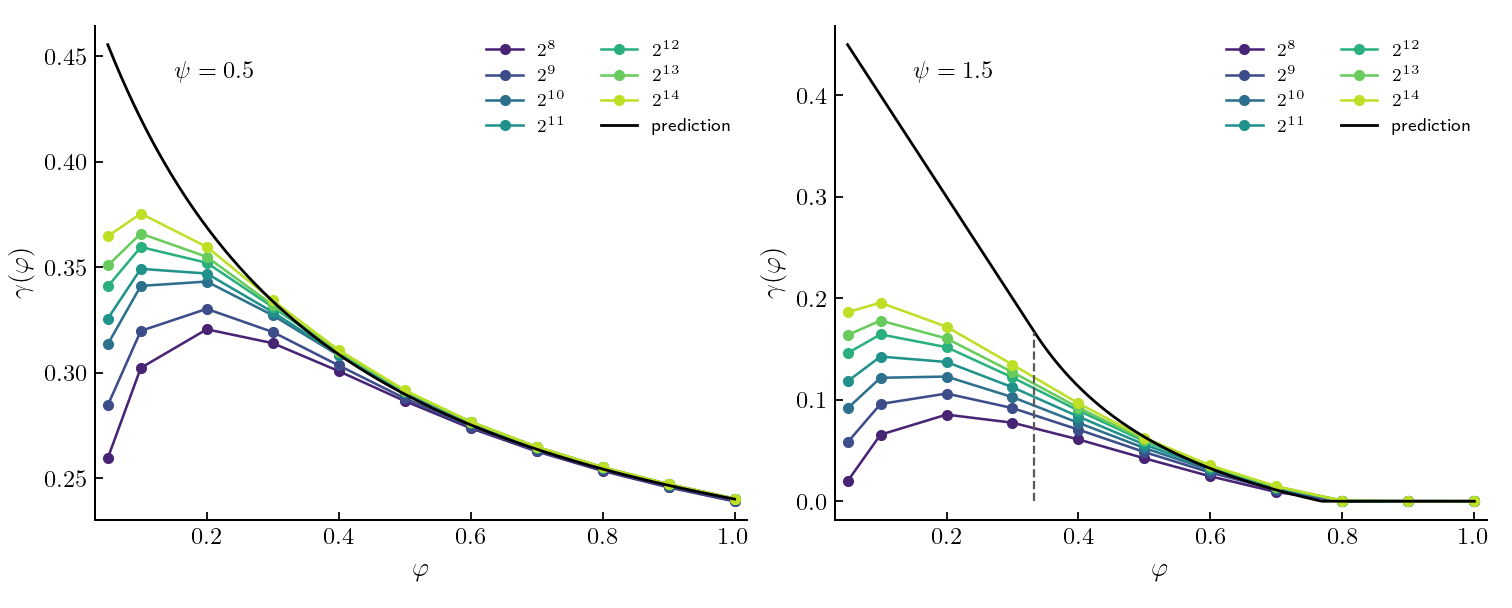}
\caption{
Numerical evaluation of the growth rate $\gamma(\varphi)$ as a function of the redistribution $\varphi$, for $r=0.5$, $m_>=1$ and several system sizes. Left: $\psi=0.5$, the system is always delocalized, the numerical estimates is consistent with the analytical prediction, although there are strong finite size effects at small $\varphi$. Right: $\psi=1.5$, there are three phases: marginal partially localized (left of the vertical dashed line), delocalized with $\gamma>0$ and with $\gamma=0$. The growth/without is predicted to occur at $\varphi_c^0\simeq 0.77$ There are strong finite size corrections at small $\varphi$, although the trend of the corrections with $N$ is consistent
}
\label{fig:numcheckcomppact}
\end{figure}

\subsection{Resulting phase diagram}

Combining the results from subsections \ref{Db} and \ref{Mplb}, we find that the overall growth rate is given by $\gamma=\max{(\gamma_{\rm pl},\gamma_{\rm deloc})}$, where $\gamma_{\rm pl}=m_>-\varphi-r$ and $\gamma_{\rm deloc}=\max{(\gamma^0-r,0)}$. The largest rate thus determines the nature of the phase.
As established earlier, there is a transition between a delocalized
phase with growth and a delocalized phase without growth occurring at $\gamma^0=r$, equivalently at $\varphi=\varphi_c^0$.
On the other hand, there is a transition between the marginal partially localized phase
 and the delocalized phase occurring when $\gamma_{\rm pl}=\gamma_{\rm deloc}$, equivalently at $\varphi=\varphi_c$. It occurs either when $\gamma_{\rm pl}=\gamma^0-r$, hence $\gamma^0+\varphi_c=m_>$ and $\varphi_c=1/G(m_>)$ or when $\gamma_{\rm pl}=0$ yielding $\varphi_c=m_>-r$. Overall, one has $\varphi_c=\min{(1/G(m_>), m_>-r)}$.
 For $\psi<1$, $G(m_>)=+\infty$ hence $\varphi_c=0$ and there is no localized/delocalized phase transition.
Note that due to the absence of a truly localized phase, one expects $\sigma_S=0$ in these phases and the results presented here should become exact as $N\to +\infty$. The predicted phase diagrams for $\psi=0.5$ and $\psi=1.5$ are illustrated in \ref{fig:phasediagcompact}.

\section{Extensions}

\subsection{Partial resetting with white noise}

Within the same self-averaging (SA) approximation, we consider the It\^o dynamics
\begin{equation}
\ddd p_i
=
\left(
a_i p_i+\frac{\varphi}{N}
\right)\ddd t
+
\sigma p_i\,\ddd W_i(t),
\qquad
a_i=m_i-\varphi-\gamma.
\label{eq:p_linear_noisy}
\end{equation}
At Poissonian times of rate $r_i$, the fraction is partially reset according to
\begin{equation}
p_i\longrightarrow f p_i+(1-f)p_0,
\qquad
0\leq f<1.
\label{eq:partial_reset_rule}
\end{equation}
The case $f=0$ corresponds to complete resetting to $p_0$.

The evolution equation for $P_i(p,t)$ reads
\begin{align}
\frac{\partial P_i(p,t)}{\partial t}
={}&
-\frac{\partial}{\partial p}
\left[
\left(
a_i p+\frac{\varphi}{N}
\right)P_i(p,t)
\right]
+
\frac{\sigma^2}{2}
\frac{\partial^2}{\partial p^2}
\left[
p^2P_i(p,t)
\right]
-r_iP_i(p,t)
\nonumber\\
&+
r_i\int \ddd p'\,
P_i(p',t)
\delta\left(
p-fp'-(1-f)p_0
\right).
\label{eq:partial_reset_noisy_fokker_planck}
\end{align}
This form is valid for all $0\leq f<1$. For $f>0$, the last term can be written as
\begin{equation}
r_i\int \ddd p'\,
P_i(p',t)
\delta\left(
p-fp'-(1-f)p_0
\right)
=
\frac{r_i}{f}
P_i\left(
\frac{p-(1-f)p_0}{f},t
\right),
\label{eq:partial_reset_gain_term}
\end{equation}
provided that the argument belongs to the support of $P_i$. For $f=0$, it instead reduces to
$r_i\delta(p-p_0)$.

In the growing phase, one may set $p_0=0$. The stationarity equation then reads
\begin{align}
0
={}&
-\frac{\ddd}{\ddd p}
\left[
\left(
a_i p+\frac{\varphi}{N}
\right)P_i(p)
\right]
+
\frac{\sigma^2}{2}
\frac{\ddd^2}{\ddd p^2}
\left[
p^2P_i(p)
\right]
-r_iP_i(p)
\nonumber\\
&+
r_i\int \ddd p'\,
P_i(p')
\delta\left(p-fp'\right).
\label{eq:partial_reset_noisy_stationary}
\end{align}
For $f>0$, the last term is $(r_i/f)P_i(p/f)$, whereas for $f=0$ it is
$r_i\delta(p)$.

Assuming the large-$p$ behavior (for $p \gg O(1/N)$)
\begin{equation}
P_i(p)\sim C_i p^{-1-\mu_i},
\label{eq:partial_reset_noisy_tail}
\end{equation}
one obtains
\begin{equation}
a_i\mu_i
+
\frac{\sigma^2}{2}\mu_i(\mu_i-1)
=
r_i\left(1-f^{\mu_i}\right).
\label{eq:partial_reset_noisy_tail_exponent}
\end{equation}
For $f=0$, this reduces to
\begin{equation}
r_i
=
a_i\mu_i
+
\frac{\sigma^2}{2}\mu_i(\mu_i-1),
\label{eq:noisy_tail_exponent_equation}
\end{equation}
whose positive solution is
\begin{equation}
\mu_i
=
\frac{
-\left(a_i-\sigma^2/2\right)
+
\sqrt{
\left(a_i-\sigma^2/2\right)^2
+
2r_i\sigma^2
}
}{
\sigma^2
}.
\label{eq:noisy_tail_exponent}
\end{equation}
Note that at fixed global parameters $\gamma$, $\varphi$, $\sigma$ and reset rate $r_i>0$, the exponent $\mu_i$ is a strictly decreasing function of the local growth rate $m_i$, so that more favorable sites have broader tails.

In the absence of resetting, $r_i=0$, one recovers the BM result
\begin{equation}
\mu_i
=
1-\frac{2a_i}{\sigma^2}
=
1-\frac{2\left(m_i-\varphi-\gamma\right)}{\sigma^2}.
\label{eq:no_reset_noisy_tail_exponent}
\end{equation}

The moment equation in the growing phase follows as
\begin{equation}
\left[
r_i\left(1-f^q\right)
-a_iq
-\frac{\sigma^2}{2}q(q-1)
\right]
\langle p_i^q\rangle
=
q\frac{\varphi}{N}
\langle p_i^{q-1}\rangle.
\label{eq:partial_reset_noisy_moment}
\end{equation}
In particular, the first moment reads
\begin{equation}
\langle p_i\rangle
=
\frac{\varphi/N}
{r_i(1-f)-a_i} =
\frac{\varphi/N}{
r_i\left[
(1-f)-\dfrac{1-f^{\mu_i}}{\mu_i}
\right]
+
\dfrac{\sigma^2}{2}(\mu_i-1)
},
\qquad
\mu_i>1.
\label{eq:partial_reset_noisy_first_moment}
\end{equation}
Note that the condition $\mu_i>1$ is equivalent to
$r_i(1-f)-a_i>0$.
The temporal noise again does not enter the first-moment equation in the It\^o convention, whereas partial resetting replaces the effective reset rate $r_i$ by
$r_i(1-f)$. Within the SA approximation the global growth rate $\gamma$ in the
delocalized phase should thus remain the same, up to the change in the effective reset rate.

For $\sigma>0$ and within the SA approximation, the support in the growing phase is $p \in [0,+\infty[$.
Note that of course the true stationary distribution $P_i(p)$ has an additional upper cutoff for $p=O(1)$, which cannot
be described within the SA approximation
- since it fails in that regime.
\\

{\bf Resulting Phase diagram}.
We now establish the corresponding phase diagram for Gaussian $m_i$ and uniform resetting $r_i=r$. In the limit of no resetting, $f \to 1^-$, one recovers the model with continuous multiplicative noise studied in Ref.~\cite{BernardMFRandomG,bernard2026fluctuations,BernardBouchaudLeDoussal2026InPreparation} for which there exist localized, delocalized and partially localized phases, which we therefore expect to exist for $0<f<1$.

{\it Delocalized phase.}
The argument is the same as in the main text. From Eq.~\eqref{eq:partial_reset_noisy_first_moment}, partial resetting replaces $r$ by $r(1-f)$ in the first-moment equation. Hence
\begin{equation}
\gamma_{\rm deloc}
=
\max\!\left[0,\mtyp-r(1-f)\right].
\label{eq:partial_reset_gamma_deloc}
\end{equation}

{\it Localized phase.}
As pointed out in \cite{BernardMFRandomG,bernard2026fluctuations,BernardBouchaudLeDoussal2026InPreparation}, localization occurs when the site with the largest growth rate, $m_1\to\mtyp+\Sigma_0$, has a probability $O(1)$ to be localized at some given time,  the localization condition is thus $\mu_1=0$. Using $(1-f^{\mu_1})/\mu_1\to-\ln f$, one gets from Eq. \eqref{eq:partial_reset_noisy_tail_exponent} that $a_1=\frac{\sigma^2}{2}-r\ln f $, which using the definition of $a_1$ in \eqref{eq:p_linear_noisy} leads to the growth rate in the localized phase
\begin{equation}
\gamma_{\rm loc}
=
\mtyp+\Sigma_0-\varphi+r\ln f-\frac{\sigma^2}{2}.
\label{eq:partial_reset_gamma_loc}
\end{equation}
which recovers the value obtained in \cite{BernardMFRandomG} for $r=0$, or equivalently $f=1$.

{\it Partially localized phase.}
From Eq.~\eqref{eq:partial_reset_noisy_tail_exponent} for the growth rate $m_i$ associated with a given $\mu_i=\mu$ gives
\begin{equation}
m(\mu)
=
\gamma+\varphi+h_f(\mu)
+\frac{\sigma^2}{2}(1-\mu).
\label{eq:partial_reset_m_of_mu}
\end{equation}
where we defined $h_f(\mu)=r\frac{1-f^\mu}{\mu}$.
For the Gaussian distribution , the $N$-dependent variance
$\Sigma_N^2=\Sigma_0^2/(2\ln N)$ gives
\begin{equation}
\rho_N(m)
\sim
\exp\left[
-\ln N\,\frac{(m-\mtyp)^2}{\Sigma_0^2}
\right]
=
N^{-(m-\mtyp)^2/\Sigma_0^2},
\end{equation}
for $m=O(1)$ and up to factors subleading in $N$. Since there are $N$ sites, the number of
sites with growth rate close to $m(\mu)$ therefore scales as
\begin{equation}
N\rho_N\!\left(m(\mu)\right)
\sim
N^{1-[m(\mu)-\mtyp]^2/\Sigma_0^2}.
\end{equation}
Moreover, the probability that this site reaches an
$O(1)$ fraction is of order $N^{-\mu}$. The contribution of these sites is therefore controlled by
\begin{equation}
\mathcal F(\mu)
=
1-\mu-
\frac{[m(\mu)-\mtyp]^2}{\Sigma_0^2}.
\end{equation}
A partially localized saddle $\mu_\star\in(0,1)$ obeys
$\mathcal F(\mu_\star)=\mathcal F'(\mu_\star)=0$. This yields
\begin{equation}
\gamma_{\rm pl}
=
\mtyp+\Sigma_0\sqrt{1-\mu_\star}
-\varphi-h_f(\mu_\star)
-\frac{\sigma^2}{2}(1-\mu_\star),
\label{eq:partial_reset_gamma_pl}
\end{equation}
where $\mu_\star$ is determined by
\begin{equation}
-h_f'(\mu_\star)+\frac{\sigma^2}{2}
=
\frac{\Sigma_0}{2\sqrt{1-\mu_\star}},
\qquad
-h_f'(\mu)
=
r\frac{1-f^\mu+\mu f^\mu\ln f}{\mu^2}.
\label{eq:partial_reset_saddle}
\end{equation}

\begin{figure}[t]
\centering
\includegraphics[width=0.9\linewidth]{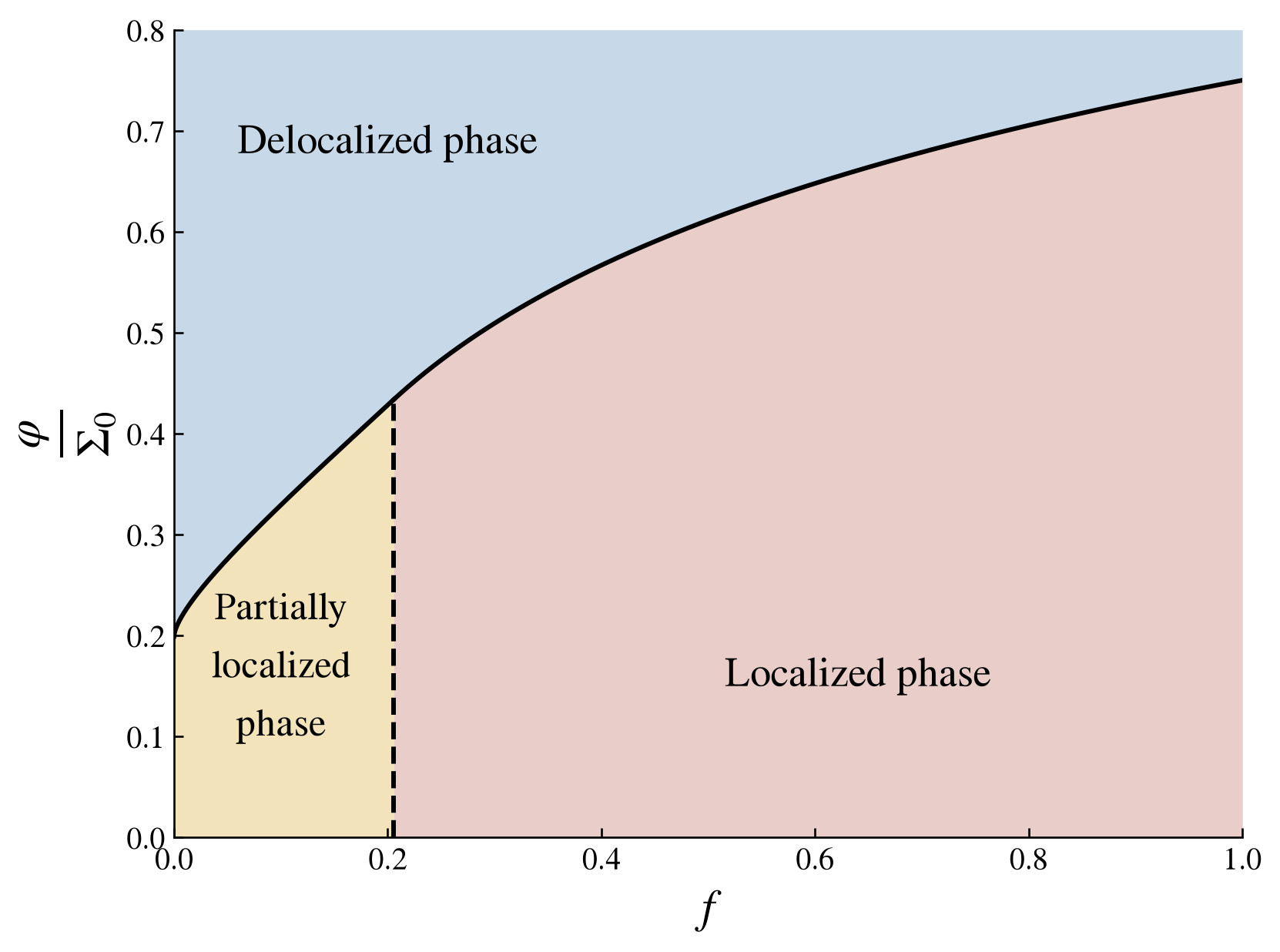}
\caption{
Phase diagram in the $(f,\varphi/\Sigma_0)$ plane for the Gaussian distribution, with $\mtyp=0$, $r/\Sigma_0=0.2$, and $\sigma^2/\Sigma_0=0.5$. Solid lines indicate where the delocalized estimate, $\gamma_{\rm deloc}=0$ here, equals either $\gamma_{\rm pl}$ or $\gamma_{\rm loc}$. The dashed line separates the partially localized and localized phases.
}
\label{fig:partial_reset_phase_diagram}
\end{figure}

The physical growth rate is the largest of the three estimates,
\begin{equation}
\gamma
=
\max\!\left(
\gamma_{\rm deloc},
\gamma_{\rm pl},
\gamma_{\rm loc}
\right),
\label{eq:partial_reset_gamma_max}
\end{equation}
and the largest estimate determines the phase. The three phases exist and a resulting phase diagram in the plane $(\varphi/\Sigma_0,f)$, with  $r/\Sigma_0=0.2$ and $\sigma^2/\Sigma_0=0.5$, is shown in Fig.~\ref{fig:partial_reset_phase_diagram}.
\\

The transition between the partially localized and localized phases occurs when the partially localized saddle reaches the edge $\mu^\star \to 0$.
Using the saddle point equation \eqref{eq:partial_reset_saddle} and $-h_f'(0)=\frac{r}{2}(\ln f)^2$, one obtains at the transition,
\begin{equation}
r(\ln f_c)^2+\sigma^2=\Sigma_0,
\end{equation}
which yields
\begin{equation}
f_c=
\exp\left[
-\sqrt{\frac{\Sigma_0-\sigma^2}{r}}
\right].
\end{equation}
This solution exists only for $\sigma^2<\Sigma_0$. For $\sigma^2\geq\Sigma_0$, the localized phase is absent, as in the model without resetting.

\subsection{Resetting to a fraction of the mean population} \label{sec:frac}

We consider two closely related variants in which the resetting value follows
the instantaneous mean population. Between resetting events, the dynamics is
unchanged.
Each site $i$ resets independently at Poissonian rate $r_i$.
We distinguish
the following two prescriptions.

In \emph{variant 1}, the reset value is a fraction $\epsilon$ of the
pre-reset mean population,
\begin{equation}
x_i(t^+)=\epsilon\,\bar x(t^-).
\label{eq:variant1_reset}
\end{equation}
The exact jump of the reset fraction is then
\begin{equation}
p_i(t^+)
=
\frac{\epsilon/N}
{1-p_i(t^-)+\epsilon/N},
\label{eq:variant1_fraction}
\end{equation}
while the total population changes according to
\begin{equation}
\frac{\bar x(t^+)}{\bar x(t^-)}
=
1-p_i(t^-)+\frac{\epsilon}{N}.
\label{eq:variant1_mean_jump}
\end{equation}

In \emph{variant 2}, the reset population is instead defined as a fraction
$\epsilon$ of the post-reset mean,
\begin{equation}
x_i(t^+)=\epsilon\,\bar x(t^+).
\label{eq:variant2_reset}
\end{equation}
In this case the fraction of the reset site is exactly
\begin{equation}
p_i(t^+)=\frac{\epsilon}{N},
\label{eq:variant2_fraction}
\end{equation}
and
\begin{equation}
\frac{\bar x(t^+)}{\bar x(t^-)}
=
\frac{1-p_i(t^-)}{1-\epsilon/N}.
\label{eq:variant2_mean_jump}
\end{equation}
The two prescriptions therefore differ when a site with an $O(1)$ fraction
is reset. They nevertheless become equivalent within the self-averaging
approximation considered below.

\paragraph{Self-averaging approximation.}
We assume that the fluctuations of the total population can be neglected and
write, at large times,
\begin{equation}
\bar x(t)\simeq x_m e^{\gamma t}.
\label{eq:mean_reset_SA}
\end{equation}
Between resets, the fractions then obey
\begin{equation}
\frac{d p_i}{dt}
=
a_i p_i+\frac{\varphi}{N},
\qquad
a_i=m_i-\varphi-\gamma.
\label{eq:mean_reset_pi_SA}
\end{equation}
Within this approximation the change of $\bar x$ during an individual reset
is neglected. Both variants therefore reduce to the same resetting rule,
\begin{equation}
p_i\longrightarrow p_0=\frac{\epsilon}{N}.
\label{eq:mean_reset_p0}
\end{equation}
The solution at a time $\tau$ after the last reset is
\begin{equation}
p_i(\tau)
=
p_i^\star+
\left(
\frac{\epsilon}{N}-p_i^\star
\right)e^{a_i\tau},
\qquad
p_i^\star=-\frac{\varphi}{Na_i}.
\label{eq:mean_reset_solution}
\end{equation}
Since $\tau$ is exponentially distributed with rate $r_i$, the stationary
distribution follows as
\begin{equation}
P_i(p)
=
|\mu_i|
\left|
\frac{\epsilon}{N}+\frac{\varphi}{Na_i}
\right|^{\mu_i}
\left|
p+\frac{\varphi}{Na_i}
\right|^{-1-\mu_i},
\qquad
\mu_i=
\frac{r_i}{m_i-\varphi-\gamma}.
\label{eq:mean_reset_stationary}
\end{equation}
In particular, for $a_i>0$ and $N^{-1}\ll p\ll1$,
\begin{equation}
P_i(p)
\sim
N^{-\mu_i}p^{-1-\mu_i}.
\label{eq:mean_reset_tail}
\end{equation}
Thus resetting to $\epsilon\bar x$ changes the $O(1)$ prefactor of the
power-law tail, but neither its exponent nor its $N$ dependence.

For sites with $\mu_i>1$ or $\mu_i<0$, whose first moment is controlled by
$p_i=O(N^{-1})$, the stationary first-moment equation gives
\begin{equation}
\langle p_i\rangle
=
\frac{\varphi+r_i\epsilon}
{N(\gamma+\varphi+r_i-m_i)}.
\label{eq:mean_reset_first_moment}
\end{equation}
The delocalized normalization condition is therefore
\begin{equation}
1\simeq
\frac{1}{N}
\sum_{\mu_i>1\ {\rm or}\ \mu_i<0}
\frac{\varphi+r_i\epsilon}
{\gamma+\varphi+r_i-m_i}.
\label{eq:mean_reset_deloc_norm}
\end{equation}

For uniform resetting rates $r_i=r$ and the generalized-Gaussian class,
most sites have $m_i\simeq m_0$, yielding at leading order
\begin{equation}
\gamma_{\rm deloc}
=
m_0-r(1-\epsilon).
\label{eq:mean_reset_gamma_deloc}
\end{equation}
Contrary to resetting to a fixed value $x_0$, there is no external population
source in the present models. The dynamics is homogeneous under a global
rescaling of all $x_i$, and $\gamma$ may therefore be negative; there is no
additional $\gamma=0$ branch.

\paragraph{Partially localized phase.}
For $0<\mu_i<1$, the first moment is controlled by the $O(1)$ cutoff and
scales as
\begin{equation}
\langle p_i\rangle\sim N^{-\mu_i}.
\label{eq:mean_reset_broad_mean}
\end{equation}
This is identical to Eqs \eqref{eq:tail_mu_beyond_sa} \eqref{eq:mean_pi_broad}, which were obtained for the resetting to a fixed value $x_0$. The saddle analysis thus yields again
\begin{equation} \label{eq:gammaplsm}
\gamma_{\rm p.l.}
=
m(u^\star)-\varphi-\frac{r}{u^\star},
\qquad
r=-m'(u^\star)(u^\star)^2.
\end{equation}
and in particular, for the generalized Gaussian case:
\begin{equation}
\gamma_{\rm p.l.}
=
m_0+\Sigma_0(1-u^\star)^{1/(2b)}
-\varphi-\frac{r}{u^\star}.
\label{eq:mean_reset_generalized_gaussian_gamma}
\end{equation}

\paragraph{Growth rate and phase diagram.}
The physical growth rate is the
larger of the delocalized and partially localized estimates,
\begin{equation}
\gamma
=
\max\left(
\gamma_{\rm deloc},
\gamma_{\rm p.l.}
\right)
=
\max\left[
m_0-r(1-\epsilon),
\gamma_{\rm p.l.}
\right].
\label{eq:mean_reset_gamma}
\end{equation}
The transition is therefore determined by
\begin{equation}
\gamma_{\rm p.l.}
=
m_0-r(1-\epsilon).
\label{eq:mean_reset_transition}
\end{equation}
Equivalently, using the saddle parametrization,
\begin{equation}
\varphi_c
=
m(u^\star)-m_0
-\frac{r}{u^\star}
+r(1-\epsilon),
\label{eq:mean_reset_phi_c}
\end{equation}
where $u^\star$ is fixed independently of $\varphi$ and $\epsilon$ by
Eq.~\eqref{eq:gammaplsm}. At fixed $r$ and
$\varphi$, the same transition may be written as
\begin{equation}
\epsilon_c
=
1+\frac{\gamma_{\rm p.l.}-m_0}{r}.
\label{eq:mean_reset_epsilon_c}
\end{equation}

{The delocalized growth rate increases with $\epsilon$ while the partially localized estimate remains the same. Increasing $\epsilon$ thus shrinks the
partially localized region. For any $\epsilon=O(1)$ and $r>0$, a reset
still sends a localized site to a fraction of order $N^{-1}$, so that
no permanently localized phase occurs within this resetting variants. Variants 1 and 2 consequently have the same self-averaging phase diagram.
A distinction might appear when one goes beyond self averaging, and in particular, considers the $O(1)$ jump
of the total population produced by the resetting of a localized site. }

\end{document}